\documentclass[prl,twocolumn,superscriptaddress]{revtex4}
\usepackage[dvips]{graphicx}
\usepackage{latexsym}
\usepackage{bm}
\newcommand{\fig}[2]{\includegraphics[width=#1,angle=-90]{#2}}
\begin{document}
\renewcommand{\ni}{{\noindent}}
\newcommand{\dprime}{{\prime\prime}}
\newcommand{\be}{\begin{equation}}
\newcommand{\ee}{\end{equation}}
\newcommand{\bea}{\begin{eqnarray}}
\newcommand{\eea}{\end{eqnarray}}
\newcommand{\la}{\langle}
\newcommand{\ra}{\rangle}
\newcommand{\dg}{\dagger}
\newcommand\lbs{\left[}
\newcommand\rbs{\right]}
\newcommand\lbr{\left(}
\newcommand\rbr{\right)}
\newcommand\f{\frac}
\newcommand\e{\epsilon}
\newcommand\ua{\uparrow}
\newcommand\da{\downarrow}
\title{BCS - BEC crossover at $T=0$: A Dynamical Mean Field Theory Approach}
\author{Arti Garg}
\affiliation{Department of Theoretical Physics,Tata Institute of
Fundamental Research, Mumbai 400 005, India}
\author{H. R. Krishnamurthy}
\affiliation{Centre for Condensed Matter Theory, Department of
Physics, Indian Institute of Science, Bangalore 560 012, India ,\\
and Condensed matter Theory Unit, JNCASR, Jakkur, Bangalore 560
064, India}
\author{Mohit Randeria}
\affiliation{Department of Theoretical Physics,Tata Institute of
Fundamental Research, Mumbai 400 005, India; \\
Department of Physics, The Ohio State University, Columbus, OH 43210}
\vspace{0.2cm}
\begin{abstract}
\vspace{0.3cm} We study the $T=0$ crossover from the BCS
superconductivity to Bose-Einstein condensation in the attractive
Hubbard Model within dynamical mean field theory(DMFT) in order to
examine the validity of Hartree-Fock-Bogoliubov (HFB) mean field
theory, usually used to describe this crossover, and to explore
physics beyond it. Quantum fluctuations are incorporated using
iterated perturbation theory as the DMFT impurity solver. We find
that these fluctuations lead to large quantitative effects in the
intermediate coupling regime leading to a reduction of both the
superconducting order parameter and the energy gap relative to the
HFB results. A qualitative change is found in the single-electron
spectral function, which now shows incoherent spectral weight for
energies larger than three times the gap, in addition to the usual
Bogoliubov quasiparticle peaks. \vspace{0.1cm} \typeout{polish
abstract}
\end{abstract}
\maketitle
\section{1. Introduction}

The problem of the crossover from BCS superconductivity to
Bose-Einstein Condensation (BEC) of composite bosons, 
where the superconducting coherence length (roughly the size of the fermion
pair binding) is respectively much larger than or much smaller
than the average inter-fermion spacing, has been a problem of
great interest from the very early stages of development of the
theory of superconductivity. It was first addressed in the very early
work of Eagles \cite{Eagles}.
In 1980 Leggett~\cite{Leggett}
showed, using a variational approach, that at zero temperature the
superconducting BCS ground state at weak coupling evolves smoothly
into a Bose condensate state of tightly bound 
"molecules" at strong coupling. Nozi\'{e}res and Schmitt-Rink~\cite{NS}
extended the analysis to lattice models and to finite temperature
and showed that the transition temperature $T_c$ between the
normal and the superconducting state evolves continuously as a
function of the magnitude of the attractive interaction between
the fermions. The discovery of high $T_c$ superconductors, which
are characterized by short coherence length comparable to (but
larger than) the inter-particle spacing, led to a resurgence of
interest in the BCS-BEC crossover. A variety of interpolation
schemes between weak and strong coupling developed using
variational methods, functional integrals, and diagrammatic
methods have been explored \cite{mohit,RDS,JRE}, and the existence of
pseudo-gap anomalies in the normal state of a short coherence
length superconductor has been established in two-dimensional
systems \cite{mohit+others,NT+MR}. Recently it has become possible
to directly realize the BCS-BEC crossover in a dilute atomic gas
of Fermions in a trap, by varying their two-body interaction
(scattering length) using a Feshbach resonance~\cite{expt}.

In this paper we analyze the BCS-BEC crossover in the attractive
Hubbard model using the dynamical mean field theory (DMFT) 
\cite{georges,jarrell} approach. Our goal is to focus on the 
intermediate coupling regime $U/t \approx 1$ where one has no 
obvious small parameter. Since the DMFT becomes exact in the 
limit of infinite dimensions \cite{georges,jarrell}, we
are, in a sense,  using the inverse coordination number of the
lattice as the small parameter. The attractive Hubbard model has
been studied recently using DMFT, but primarily in the normal
phase~\cite{DMFT} to analyze pair formation above $T_c$ and
related phenomena. We focus here on the superconducting phase at
zero temperature, in part because the DMFT method has been much
less explored in broken symmetry phases.

The remainder of the paper is organized as follows. In Section 2,
we define the model and review the Hartree-Fock-Bogoliubov (HFB)
mean field theory. In Section 3, we briefly summarize the DMFT
approach in the superconducting (SC) state and then describe the
specific implementation of DMFT which we use, namely the iterated
perturbation theory (IPT), in Section 4. In Section 5 we present
our results for the chemical potential, energy gap, SC order
parameter, density of states, spectral function, occupation
probability and superfluid stiffness. We discuss how each of these
evolves from the weak coupling BCS limit to the strong couping BEC
limit, and to what extent the quantum fluctuations included in the
DMFT implemented using IPT modify the results relative to HFB mean
field theory.

\section{2. Model}
We use the simplest lattice model which exhibits the BCS-BEC
crossover, defined by the Hamiltonian 
\be H = - t
\sum_{ij,\sigma}c^{\dagger}_{i\sigma}c_{j\sigma}
-|U|\sum_{i}n_{i\ua}n_{i\da}-\mu\sum_{i}n_{i} \label{hamiltonian}
\ee 
The first term describes the kinetic energy of fermions with
nearest neighbor hopping $t$, the on-site attractive interaction 
($-|U|$) induces s-wave, singlet pairing and leads to a
superconducting ground state for all $n \ne 1$, with the chemical
potential $\mu$ determining the filling factor $n$. (We will not
study the system with $n = 1$ for which superconducting and
charge-density wave orders coexist).

The simplest mean-field description of this system uses the
Hartree-Fock-Bogoliubov (HFB) theory leading to the following
self-consistent equations for the ``gap'' $\Delta$ and $\mu$ at a
temperature $T \equiv 1/(k_B \beta$): \be \f{1}{|U|}= \sum_{k}
\f{\tanh(\beta E_{k}/2)}{2E_{k}} \label{HF_delta} \ee and \be
n=2\sum_{k} \lbs 1-\f{\xi_{k}}{E_{k}}\tanh\lbr\f{\beta
E_{k}}{2}\rbr \rbs. \label{HF-noeqn} \ee We use standard notation
where $\epsilon_{k}$ is the band dispersion for the fermions and
$E_{k}=\sqrt{\Delta^{2}+\xi_{k}^{2}}$ with $\xi_{k} \equiv
\epsilon_{k}-\mu-|U|n/2$. 
As is well known \cite{mohit,RDS,JRE,mohit+others3}
at $T=0$ the solution of these equations leads to a BCS
superconductor in the weak coupling $|U|/t \ll 1$ limit, to a BEC
of hard core bosons in the opposite extreme $|U|/t \gg 1$, and
interpolates smoothly in between. However, the finite temperature
solutions of these equations for $|U|/t \gg 1$ yields a $T_c \sim
|U|$, not the BEC transition temperature scale
expected to be of order $t^2/|U|$, and the HFB approach therefore
does not constitute an interpolating approximation at finite $T$
\cite{mohit}.

The DMFT is one of the simplest schemes that has the potential to
overcome some of these limitations of simple HFB theory. As we
discuss in the following sections, the lattice dependence of
quantities that arise in the DMFT is not via the momentum $k$ but
only via the band dispersion $\epsilon_{k}$, and hence we can make
the replacement \be \sum_{k} \to \int d\epsilon \rho(\epsilon),
\ee where $\rho(\epsilon)$ is the (bare) band density of
states(DOS). The implementation of the DMFT is often simplest on a
Bethe lattice with a large coordination number $z~\rightarrow
\infty$, for which \be \rho(\epsilon) =
{\sqrt{4t^{2}-\epsilon^{2}}\over{2\pi
t^{2}}}\theta(2t-|\epsilon|). \label{bethe_dos} \ee  if the bare
hopping is normalized as \be t\rightarrow\f{t}{\sqrt{z}}
\label{t_scaling} \ee

\begin{figure}
\begin{center}
\vskip -2.0cm
\hspace*{-3.5cm}
\centerline{\fig{2.8in}{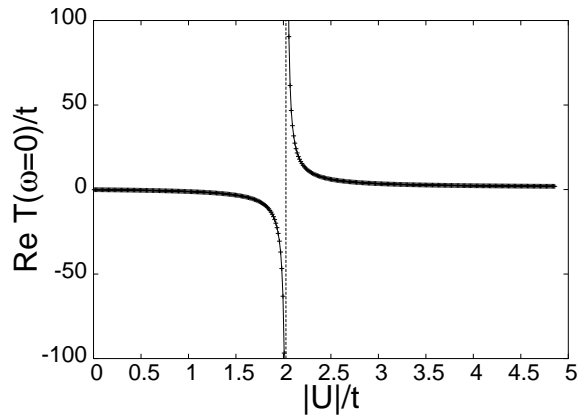}}
\vskip 0.5cm \caption{The real part of the T-matrix $T(\omega \to
0)$ for the two-body problem is plotted as a function of the
attraction $|U|/t$ between two fermions in an otherwise empty
Bethe lattice of infinite connectivity. $|U|/t=2$ is the threshold
for the formation of bound state in vacuum.} \label{Tmatrix}
\end{center}
\vskip-6mm
\end{figure}

We conclude this section by calculating the effective two-body
interaction or the low energy scattering amplitude. This will give 
us a clear idea about the regime of $|U|/t$ where we expect the 
corrections to HFB at $T=0$ to be the most severe, and it will also
emphasize the similarity between the continuum Fermi gases
often studied theoretically (and now experimentally) and the lattice
model studied in this paper. The low energy scattering
is described by the real part of the T-matrix ${\rm
Re}T(\omega \to 0)$ for the two-body problem in vacuum, i.e., for two
fermions in an otherwise empty lattice. This is the analog of the
well known three-dimensional ``scattering length'' for the case of
the Bethe lattice studied in this paper. 
As shown in Fig.~\ref{Tmatrix}, for $|U|/t < 2$ the
attractive interaction is not sufficient to cause a two-body
bound state in vacuum, and $|U|/t = 2$ is the threshold for bound state
formation at which the scattering amplitude diverges \cite{Sakurai}.
We also note that at $|U|/t=2$ the effective interaction diverges, i.e., 
one reaches the unitary limit, even though bare $|U|$ is in the 
intermediate coupling regime. We expect that the deviations from the HFB 
theory will be maximal in the vicinity of $|U|/t = 2$ where the
system is efectively very strongly interacting.

\section{3. Dynamical mean field theory}
To explore the intermediate coupling regime we use the dynamical
mean field theory (DMFT) approach \cite{georges,jarrell}, which reduces a
lattice problem with many degrees of freedom to an effective
single-site problem by "integrating out" all the fermionic degrees
of freedom except those at one site -- the ``impurity site'' --
and retaining the effects of this only in the form of a
self-consistently determined bath with which the "impurity site"
hybridizes. This retains non-trivial local quantum fluctuations
missing in conventional mean field theories and the description
can be shown to be exact in the limit of large dimensionality.
Since there are many excellent reviews of DMFT we will only
outline the elements of the technique in order to introduce our
notation and to indicate the changes in the standard formalism
necessitated by the presence of the superconducting long range
order.

To take the superconducting order (with singlet pairing) into
account we use the Nambu formalism with the spinors
$\Psi^{\dagger}_{k} \equiv (c^{\dagger}_{k\ua}, c_{-k\da})$ and
the matrix Green's function \bea
\hat{G}(k,\tau) \equiv -\langle T_{\tau} \Psi(k,\tau) \Psi^{\dagger}(k,0)\rangle \nonumber \\
\mbox{~~~~~~~~}= \lbr \begin{array} {cc} G(k,\tau) & F(k,\tau) \\
F^{\dagger}(k,\tau) & -G(-k,-\tau) \end{array} \rbr \eea where
$F(k,\tau) \equiv -\langle
T_{\tau}c_{k\ua}(\tau)c_{-k\da}(0)\rangle$ satisfies $
F(-k,-\tau)=F(k,\tau)$. We will denote all Nambu matrices by a
`hat' on top. In this formalism the interaction effects are
described in terms of the self energy matrix \be
\hat{\Sigma}(k,i\omega_{n})=\lbr
\begin{array}{cc} \Sigma(k,i\omega_{n}) & S(k,i\omega_{n}) \\
S^{\star}(k,-i\omega_{n}) & -\Sigma^{\star}(k,i\omega_{n})
\end{array}\rbr \ee where $\omega_{n}=(2n+1)\pi/\beta$ are
fermionic Matsubara frequencies.

In the limit of infinite dimensions it can be shown that the self
energy is {\em purely local}, i.e., is k independent (see ref.
\onlinecite{georges}), so that $\hat{\Sigma} =
\hat{\Sigma}(i\omega_{n})$ . Furthermore, the SC order parameter
can be chosen to be real in a uniform system, which implies that
$S(i\omega_{n})=S^{\star}(-i\omega_{n})$. Hence using the Dyson
equation the full Green's function for the lattice can be written
as \be \hat{G}^{-1}(k,i\omega_{n}) = \lbr  \begin{array}{cc}
i\omega_{n}+\mu-\epsilon_{k} & 0
\\ 0 & i\omega_{n}-\mu+\epsilon_{k} \end{array}\rbr - \hat{\Sigma}(i\omega_{n})
\label{lat-GF}\ee
$$ = \lbr  \begin{array}{cc} i\omega_{n}+\mu-\epsilon_{k}-\Sigma(i\omega_{n})
& -S(i\omega_{n}) \\ -S(i\omega_{n}) &
i\omega_{n}-\mu+\epsilon_{k}+ \Sigma(-i\omega_{n})
\end{array}\rbr $$ Thus, in DMFT the k-dependence in
$\hat{G}^{-1}(k,i\omega_{n})$ enters only via the dispersion
$\epsilon_{k}$.

The local self energy is itself obtained from an effective single
site problem which can be regarded as arising from integrating out
fermionic variables on all sites except one. The effective action
for this single site problem within the DMFT approximation is
given by
$$ S_{eff}=-\int_{0}^{\beta}d \tau \int_{0}^{\beta}d \tau^{\prime}
\Psi^{\dagger}(\tau)\hat{\mathcal{G}}^{-1}(\tau
-\tau^{\prime})\Psi(\tau^{\prime})$$ \be
\mbox{~~~~~~~~~~~~~~~~~~~~~~~~~~~}- |U| \int_{0}^{\beta} d\tau
n_{\da}(\tau)n_{\ua}(\tau). \label{impurity-action} \ee Here the
host (Matrix) Green's function $\hat{\mathcal{G}}$ is {\em not}
the non-interacting local Green's function, as it includes the
effects of the fermionic degrees at other sites which have been
integrated out {\it in the presence of interactions}, i.e., it
includes (local) self energy corrections at all these other sites,
and needs to be determined by a triangle of self-consistency
relations as described below.

The first of these relations comes from the requirement that the
"impurity" Green's function for the single site problem should be
the same as the local Green's function of the lattice, so that \be
\hat{G}(i\omega_{n})=\sum_{k}\hat{G}(k,i\omega_{n}). \ee This
gives the diagonal and off-diagonal components of the impurity
Green's function as \be G(i\omega_{n}) =
\int_{-\infty}^{\infty}d\epsilon
~\rho(\epsilon)\frac{\zeta_{1}-\epsilon}{(\zeta_{1}-\epsilon)(\zeta_{2}-\epsilon)
+S^{2}(i\omega_{n})}~, \ee and \be F(i\omega_{n}) =
S(i\omega_{n})\int_{-\infty}^{\infty}d\epsilon
~\rho(\epsilon)\frac{1}{(\zeta_{1}-\epsilon)(\zeta_{2}-\epsilon)
+S^{2}(i\omega_{n})} \label{full_G} \ee Here $\zeta_{1} \equiv
i\omega_{n}+\mu-\Sigma(i\omega_{n})$ and $\zeta_{2} \equiv
i\omega_{n}+\mu+\Sigma(-i\omega_{n})$. For the case of the
semi-circular DOS of eq.~(\ref{bethe_dos}) the integrals can be
evaluated in closed form as \be G(i\omega_{n})  = \f{2
\zeta_{1}}{x_{1}-x_{2}}\lbs \f{1}{x_{1}+\sqrt{x_{1}^{2}-4t^{2}}} -
\f{1}{x_{2}+\sqrt{x_{2}^{2}-4t^{2}}} \rbs \label{G_bethe} \ee \be
F(i\omega_{n}) = \f{S(i\omega_{n})}{x_{1}-x_{2}}\lbs
\f{1}{x_{1}+\sqrt{x_{1}^{2}-4t^{2}}} -
\f{1}{x_{2}+\sqrt{x_{2}^{2}-4t^{2}}} \rbs \label{F_bethe} \ee with
$x_{1,2} \equiv \zeta_{2}-\zeta_{1}\pm
\sqrt{(\zeta_{2}+\zeta_{1})^{2}-4S^{2}(i\omega_{n})}/2$ .

The second relation comes from the Dyson equation connecting the
full Green's function $\hat{G}$ at the impurity site, the host
Green's function $\hat{\mathcal{G}}$ and the self-energy
$\hat{\Sigma}$, typically used in reverse, in the form \be
\hat{\mathcal{G}}^{-1}(i\omega_{n}) = \hat{G}^{-1}(i\omega_{n}) +
\hat{\Sigma}(i\omega_{n}). \label{dyson} \ee

The final relation comes from the solution for the self energy of
the impurity problem defined by (\ref{impurity-action}), i.e., the
determination of  \be \hat{\Sigma}(i\omega_{n}) =
\hat{\Sigma}[\hat{\mathcal{G}}(i\omega_{n})] \label{selfenergy}
\ee from a knowledge of the host Green's function. This is the
task of the ``impurity solver'', and is typically the hardest step
in the triangle of self consistency. In this paper, we use {\em
iterated perturbation theory} (suitably extended to deal with the
broken symmetry associated with the superconducting ground state
as described in the following section) as the impurity solver.

\section{4. Iterated Perturbation Theory}

We adapt the iterated perturbation theory (IPT), originally
developed for the paramagnetic phase of the repulsive Hubbard
model \cite{georges,kk}, to the SC phase of the attractive Hubbard
model. IPT is an approximate technique which is much simpler than
the more accurate but elaborate alternate methods such as quantum
Monte Carlo \cite{QMC}, exact diagonalization \cite{ED}, numerical 
renormalization group \cite{NRG}, local moment approximation \cite{LMA} etc.
IPT gives semi-analytical results which can be directly and easily
continued to the real frequency domain. It has been well studied in the 
context of the DMFT of the Mott transitions in the repulsive Hubbard model 
\cite{georges,kk} where it gives results in complete qualitative agreements 
with the more accurate methods mentioned above, and only quantitative 
disagreement typically no more than 10-20 $\%$ in the transition 
temperatures and critical values of $U/W$; see, e.g., the comparison of 
results obtained using different impurity solvers by Bulla et al. \cite{NRG}. 
One can reasonably expect similar levels of qualitative and quantitative 
correctness in the present context in general. If qualitative changes are 
likely,  this is commented on at appropriate places in the paper. 

In the form in which we use it here \cite{lt-note}, IPT rests on the following ansatz for
the self energy as a functional of $\hat{{\mathcal{G}}}$ : 
\be
\hat{\Sigma}_{IPT}(\omega^{+})=\hat{\Sigma}_{HFB}+
\hat{A}\hat{\Sigma}^{(2)}(\omega^{+}). \label{ipt_ansatz} \ee 
Here
$\hat{\Sigma}_{HFB}$ is the Hartree-Fock-Bogoliubov (HFB) self
energy as in eq.~\ref{se_hfb} (see note \cite{ipt-hfb-note}),
$\hat{\Sigma}^{(2)}$ is the second order perturbation theory
result (in powers of $|U|$) {\it but calculated in terms of the
Hartree-corrected host Green's function} (see
eq.~\ref{hartree_corrected_host_se}), and $\hat{A}$ is to be
determined as described below (see eq.~\ref{parA}). All the
calculations we report and discuss in this paper are done at
$T=0$, and we work directly in real frequency $\omega^{+} = \omega
+ i0^{+}$.

The IPT ansatz is constructed so that \cite{lt-note} it
\begin{itemize}
\item reproduces the leading order terms for the self energy in
the weak coupling limit $|U|/t<<1$, \item is exact in the atomic
limit $t/|U|=0$, and \item reproduces the leading order terms for
the self energy in the large $\omega$ limit for all $|U|/t$, which
ensures that some exact sum rules are satisfied.
\end{itemize}
Thus IPT is expected to provide a reasonable interpolating scheme
between the weak and strong coupling limits.

The HFB self energy is given by \be \hat{\Sigma}_{HF} =
-|U|\f{n}{2} \hat{\tau}_{z}-\Delta \hat{\tau}_{x} \label{se_hfb}
\ee Here $\hat{\tau}_{z}$ and $\hat{\tau}_{x}$ are Pauli matrices
in Nambu space. The filling factor $n = \sum_{\sigma}\langle
c^{\dagger}_{\sigma}c_{\sigma}\rangle$ and $\Delta = |U|\Phi =
|U|\langle c_{\downarrow}c_{\uparrow}\rangle $ with $\Phi$ being
the superconducting order parameter, are obtained from the full
Green's function within IPT as \be n=-\f{2}{\pi}
\int_{-\infty}^{0} \mbox{Im~} G(\omega^{+}) d \omega
\label{no_eqn} \ee \be \Delta=-\f{|U|}{\pi} \int_{-\infty}^{0}
\mbox{Im~} F(\omega^{+}) d \omega \label{delta} \ee

The diagonal and off-diagonal components of the second order self
energy $\hat{\Sigma}^{(2)}$ are given by \be \Sigma^{(2)}(t)=
-U^{2}\lbr \tilde{\mathcal{G}}_{11}(t)
\tilde{\mathcal{G}}_{22}(-t)\tilde{\mathcal{G}}_{22}(t)-
\tilde{\mathcal{F}_{0}}^{\dagger}(t)\tilde{\mathcal{G}}_{22}(-t)
\tilde{\mathcal{F}_{0}}(t)\rbr \label{sigma2} \ee and \be
S^{(2)}(t)= -U^{2}\lbr
\tilde{\mathcal{F}_{0}}(t)\tilde{\mathcal{F}_{0}}(-t)
\tilde{\mathcal{F}_{0}}^{\dagger}(t)-\tilde{\mathcal{G}}_{11}(t)
\tilde{\mathcal{F}_{0}}(-t)\tilde{\mathcal{G}}_{22}(t)\rbr
\label{anomalous_sigma2} \ee
Here $\tilde{\mathcal{G}}_{11}$, $\tilde{\mathcal{G}}_{22}$, $\tilde{\mathcal{F}}_{0}^{\dagger}$ and
$\tilde{\mathcal{F}_{0}}$ are components of the Hartree corrected
Host Green's function matrix 
\be \lbr  \begin{array}{cc}
\tilde{\mathcal{G}}_{11}(\omega) & \tilde{\mathcal{F}_{0}}(\omega)
\\ \tilde{\mathcal{F}}_{0}^{\dagger}(\omega) &
\tilde{\mathcal{G}}_{22}(\omega)\end{array}\rbr ^{-1} =\lbr
\begin{array}{cc} \mathcal{G}_{0}(\omega) &
\mathcal{F}_{0}(\omega) \\ \mathcal{F}_{0}(\omega) &
-\mathcal{G}_{0}^{\star}(-\omega)\end{array}\rbr ^{-1}
-\hat{\Sigma}_{HFB} \label{hartree_corrected_host_se} \ee 
The
subscript $0$ has been added in order to distinguish the
components of the host green functions that arise in the specific
context of the IPT approximation to the impurity problem.

Each of the terms in (\ref{sigma2}) and (\ref{anomalous_sigma2})
is the product of three factors of the form \be
H(t)=h_{1}(t)h_{2}(-t)h_{3}(t) \ee where each $h_{i}$ is either
$\tilde{\mathcal{G}}_{11},\tilde{\mathcal{G}}_{22},\tilde{\mathcal{F}}_{0}^{\dagger}$
or $\tilde{\mathcal{F}_{0}}$. Using the spectral representation
for each Green's function we obtain for the Fourier transform, \be
H(\omega^{+}) = -\int_{-\infty}^{\infty}
\prod_{i=1}^{3}d\epsilon_{i} \tilde{\rho}_{i}(\epsilon_{i})
\f{N(\epsilon_{1},\epsilon_{2},\epsilon_{3})}
{\omega^{+}-\epsilon_{1}+\epsilon_{2}-\epsilon_{3}} \ee where
$\tilde{\rho}_{i}(\epsilon_{i})=-\mbox{Im}[h_{i}(\epsilon_{i}^{+})]/\pi$
and $N(\epsilon_{1},\epsilon_{2},\epsilon_{3})$ is a thermal
factor \be
N(\epsilon_{1},\epsilon_{2},\epsilon_{3})=f(\epsilon_{1})f(-\epsilon_{2})
f(\epsilon_{3})+f(-\epsilon_{1})f(\epsilon_{2})f(-\epsilon_{3})
\ee involving the Fermi function $f(\epsilon) \equiv
1/[1+\exp(\beta\epsilon)] = 1- f(-\epsilon)$.

The matrix $\hat{A}$ in (\ref{ipt_ansatz}) is fixed by demanding
that $\hat{\Sigma}_{IPT}(\omega^{+})$ is "exact" in the large
$\omega$ limit up to order $1/\omega$. As shown in Appendix A, we
find $\hat{A}$ to be proportional to the identity matrix
$\hat{\tau}_{0}$ in Nambu space and given by \be \hat{A}=\lbs
\f{U^{2}n_{0}}{2}\lbr 1-\f{n_{0}}{2}\rbr-\Delta_{0}^{2}\rbs
^{-1}\lbs \ \f{U^{2}n}{2}\lbr
1-\f{n}{2}\rbr-\Delta^{2}\rbs\hat{\tau}_{0} \label{parA} \ee where $\hat{\tau}_{0}$ is the identity matrix. Here
$n_{0}$ and $\Delta_{0}$ denote fictitious ``filling factor'' and
``gap function'' values evaluated for the Hartree corrected Host
Green's function, i.e., \be n_{0}=-2/\pi \int
_{-\infty}^{0}\mbox{Im}\tilde{\mathcal{G}}_{11}(\omega^{+})d
\omega \ee and \be \Delta_{0}=-|U|/\pi \int_{-\infty}^{0}
\mbox{Im}\tilde{\mathcal{F}_{0}}(\omega)d\omega. \ee

In the atomic limit, as discussed in Appendix B, we find that the
second order self energy vanishes. Thus the IPT self energy for
$t/U = 0$ is simply the HFB self energy. We show in Appendix B
that the HFB result is exact at zero temperature in the broken
symmetry phase for $t/U=0$, and thus our ansatz for the self
energy is exact in the atomic limit.

\section{5. Results}
We have solved the DMFT equations within the IPT approximation as
follows. For a given $|U|/t$ and $n$, we start with a guess for
the self energy $\hat{\Sigma}(\omega^{+})$ and the chemical
potential $\mu$ . With this self energy as input, we compute the
full local Green's function $\hat{G}(\omega^{+})$ using analytically continued form of 
eqs.~(\ref{G_bethe}) and (\ref{F_bethe}) at $T=0$. Then we use the Dyson
equation (\ref{dyson}) to determine the host Green's function
$\hat{\mathcal{G}}_{}$. Next we use the IPT ansatz (\ref{ipt_ansatz}-\ref{hartree_corrected_host_se})
to determine the (new) self energy in terms of
$\hat{\mathcal{G}}_{}$, using new values of parameters
$\Delta,\Delta_{0}$ and $n_{0}$. Finally we obtain the new
chemical potential  by solving the filling constraint equation
(\ref{no_eqn}) using the Broyden method~\cite{broyden}. We then
iterate the whole procedure until a self-consistent solution is
reached, i.e., convergence in the self energy matrix is achieved.

Within this self-consistency loop the evaluation of $n$ and
$\Delta$ using eq.~(\ref{no_eqn}) and (\ref{delta}) (and similarly
for $n_{0}$ and $\Delta_{0}$) involves integrals with singular
integrands: the functions Im G($\omega^{+}$) and Im
F($\omega^{+}$) have square root singularity at a gap edge $\omega
= E_g$ which is not a priori known. We fit these functions in a
small neighborhood of the gap edge to the form
$K/\sqrt{\omega-E_g}$, where the fits determine the gap in the
spectrum $E_g$. Then the singular part of integral near the gap
edge is easily evaluated analytically and the part away from the
singularity evaluated numerically using Gaussian quadrature.

All of the results reported in this paper have been obtained at a
fixed density of $n=0.5$ (quarter filling) in order to avoid
special features that arise at half-filling. (At half-filling,
corresponding to $n = 1$,  charge density wave order becomes
degenerate with SC order and the Hamiltonian has SU(2) symmetry,
and is in fact isomorphic to the repulsive Hubbard model which has
been well studied within DMFT.) We expect similar results for $n
\ne 1$.

\begin{figure}
\begin{center}
\vskip -2cm \hspace*{-2.5cm} \centerline{\fig{2.8in}{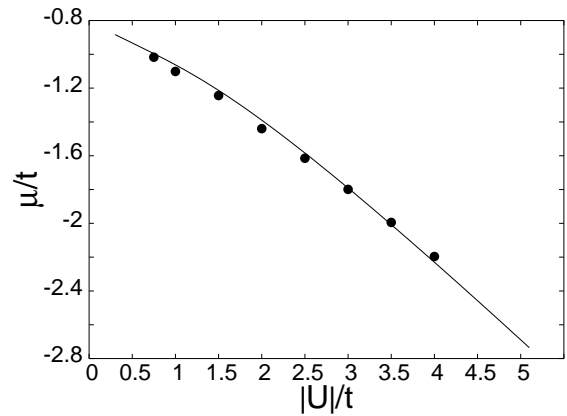}}
\vskip 0.5cm \caption{The chemical potential $\mu/t$ as a function
of $|U|/t$ for $n=0.5$ and $T=0$ within IPT (filled circles) and
HFB theory (full line).}
\label{mu}
\end{center}
\vskip-6mm
\end{figure}

{\bf Chemical potential:}  Fig.~\ref{mu} shows the chemical
potential $\mu$ tuned to obtain $n=0.5$ at $T=0$. We see that it
decreases monotonically as a function of $|U|/t$ and the system
becomes non-degenerate with increasing attraction between the
fermions. For $|U| > 3.5t$ the chemical potential goes below the
bottom of the band.

\begin{figure}
\begin{center}
\vskip 0.1cm
\hspace*{0.5mm}
\includegraphics[width=2.8in,angle=0]{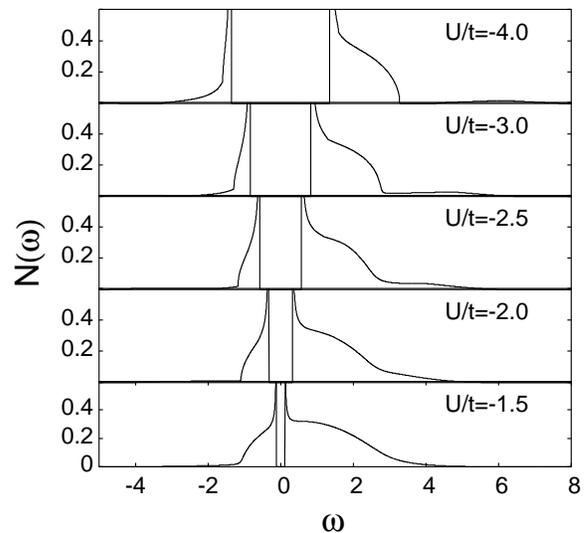}
\vskip 0.5cm \caption{Single particle density of states (per unit energy per unit area) for
$n=0.5$ and $T=0$ within IPT for Bethe lattice of infinite
connectivity.} \label{dos}
\end{center}
\vskip-6mm
\end{figure}
\begin{figure}
\begin{center}
\vskip-2cm \hspace*{-3.5cm} \centerline{\fig{2.8in}{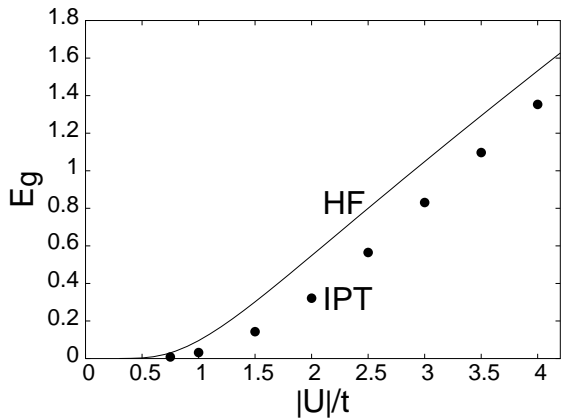}}
\vskip 0.5cm \caption{The spectral gap in the single particle
density of states for $n=0.5$ and $T=0$ as a function of $|U|/t$
within IPT (filled circles) and HFB theory (full line). Note
that the spectral gap within IPT is suppressed as compared to that
obtained from HFB theory.} \label{gap}
\end{center}
\vskip-6mm
\end{figure}

{\bf Density of states:} The single particle density of states
(DOS) N($\omega$) \be N(\omega) = -\f{1}{\pi} \mbox{Im
G}(\omega^{+}) \ee is plotted in Fig.~\ref{dos} for various values
of $|U|/t$.
We observe a spectral gap ($E_g$) in the
single particle DOS , which increases with $U|/t$ as shown in
Fig.~\ref{gap}, where the gap as obtained within the HFB theory is
also shown for comparison.

For weak coupling the HFB spectral gap has the form $t\exp(-\pi
t/2|U|)$, characteristic of BCS theory, while for large attraction
it approaches to the binding energy of the composite bosons being
proportional to $ |U|/2$. The differences of the DMFT result for
the energy gap from the simple HFB estimates will be discussed
below. Near the gap edge the DOS has a square root singularity
characteristic of a s-wave superconductor. But the DOS far from
the gap edge does not simply look like the non-interacting
semi-circular DOS of the Bethe lattice (as would be the case in
weak coupling BCS theory). The structure at larger energy values
comes from the $\omega$ dependence of the self energy, as we
discuss below.

\begin{figure}
\begin{center}
\vskip -2cm \hspace*{-3.5cm} \centerline{\fig{2.8in}{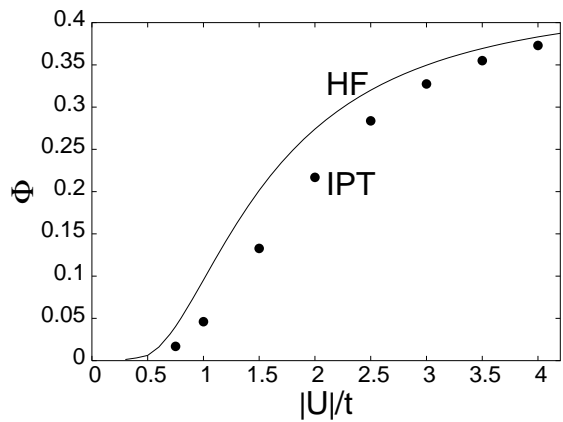}}
\vskip 0.5cm \caption{The superconducting order parameter $\Phi$
for $n=0.5$ and $T=0$ within IPT (filled circles) and HFB theory
(full line). Note that the order parameter within IPT is
suppressed as compared to that obtained from HFB theory.}
\label{op}
\end{center}
\vskip-6mm
\end{figure}

\begin{figure}
\begin{center}
\vskip -2cm \hspace*{-2.5cm}
\centerline{\fig{2.8in}{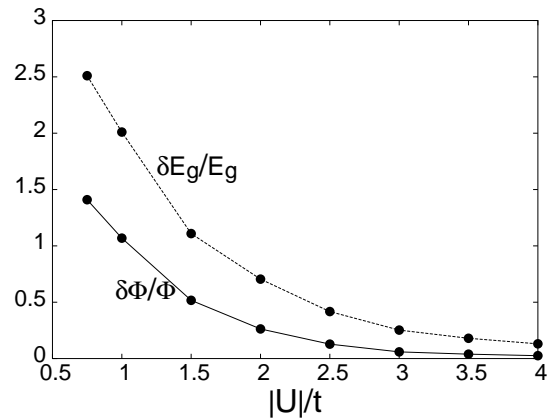}} \vskip 0.5cm
\caption{$\delta \Phi=\Phi_{HFB}-\Phi_{IPT}$  where $\Phi_{HFB}$
is the SC order parameter within HFB theory and $\Phi_{IPT}$ is
the same within IPT. $\delta E_g=(E_g)_{HFB}-(E_g)_{IPT}$ where
$(E_g)_{HFB}$ is the spectral gap within HFB theory and
$(E_g)_{IPT}$ is the same within IPT.} \label{deviation}
\end{center}
\vskip-6mm
\end{figure}

{\bf Order parameter and energy gap:}
The superconducting order-parameter is calculated using
\[\Phi=\langle c_{\da}c_{\ua}\rangle=-1/\pi \int_{-\infty}^{0} \mbox{Im} F(\omega^{+}) d\omega\]
and plotted in Fig.~\ref{op}. We see that, as expected, the
quantum fluctuations included in DMFT suppress the order parameter
below its HFB mean-field value. The effect of quantum fluctuations
in the intermediate coupling regime can be seen more clearly in
Fig.~\ref{deviation} where we plot the fractional deviation of
DMFT order parameter and the energy gap from their corresponding
HFB values. For small to intermediate values of the coupling $U
\stackrel {\sim} {<} t$ the quantitative differences are quite
large with the HFB results being larger than the DMFT ones by more
than $100\%$.

{\bf Spectral function:}

Another important quantity of interest is the one-particle
spectral function \be A(\epsilon,\omega) = - {1 \over \pi} {\rm
Im}G(\epsilon,\omega^{+}) \label{a_def} \ee where $G$ is the
``11'' component of the Nambu Matrix Green's function for the
lattice obtained by inverting eqn. \ref{lat-GF}, and is given by
\be G(\epsilon,\omega^{+})=\f{\omega+\epsilon-\mu+
\Sigma^{*}(-\omega^{+})}{D(\epsilon,\omega)} \label{g_def} \ee
with \be D(\epsilon,\omega) =
\mbox{~~~~~~~~~~~~~~~~~~~~~~~~~~~~~~~~~~~~~~~~~~~~~~~~~~~~} \ee
$$
\lbs\omega+\epsilon-\mu+\Sigma^{*}(-\omega^{+})\rbs \lbs\omega-
\epsilon+\mu-\Sigma(\omega^{+})\rbs-S^{2}(\omega^{+})
$$
Since we are working within the DMFT framework, we have traded the
${\bf k}$ label for the energy label $\epsilon$.

\begin{figure}
\begin{center}
\vskip-2.5cm \hspace*{-2.5cm}
\centerline{\fig{3.2in}{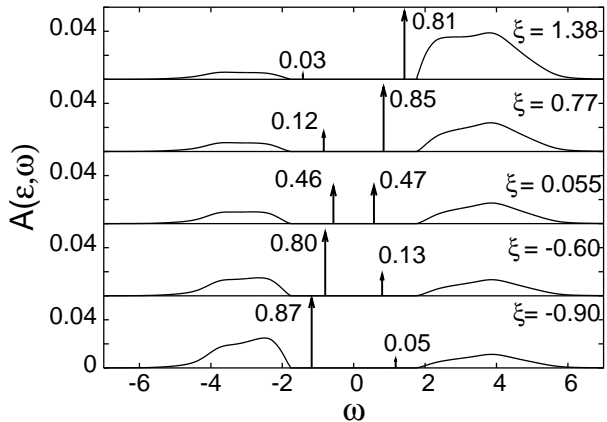}} \vskip 0.5cm \caption{The
spectral function $A(\epsilon,\omega)$ for $n=0.5$, $T=0$ and
$|U|=2.5t$ as a function of $\omega$ for various values of
$\xi=\epsilon-\mu+\Sigma^{\prime}(\omega=E)$. Note that
$A(\epsilon,\omega)$ not only has coherent delta function peaks
(which are shown by arrows with the corresponding weights) but
also has broad incoherent parts which start appearing for $\omega
\ge 3E_g$. For $|U|=2.5t$, $E_g=0.57t$.} \label{akw}
\end{center}
\vskip-6mm
\end{figure}

Quite generally, we expect that the spectral function will be of
the form \be A(\epsilon,\omega)=Z_{+}(\epsilon)\delta(w-E)+Z_{-}
(\epsilon)\delta(w+E)+ A_{inc}(\epsilon,\omega) \ee where
$Z_{\pm}(\epsilon)$ are the coherent spectral weights in the
Bogoliubov quasiparticle/quasihole excitation poles at energies
$\pm E(\epsilon)$, and $A_{inc}$ is the incoherent part of the
spectral function. We recall that in simple BCS-HFB mean field
theory $Z_{\pm} = (1 \pm \xi/E)/2$ where $\xi=\epsilon-\mu-|U|n/2$ and $E =
\sqrt{\xi^2 + \Delta^2}$ and $A_{inc} = 0$. In contrast, as
shown in Fig.~\ref{akw} the DMFT result for $A(\epsilon,\omega)$
not only has sharp delta function peaks at $\pm E$ corresponding
to the Bogoliubov excitations but also has a broad incoherent
part. The weight in the coherent excitations $Z_+ + Z_{-} < 1$ and
the deficit from unity is contained in $A_{inc}$. All of this is a
consequence of the frequency dependent self-energy as shown below.

\begin{figure}
\begin{center}
\vskip-2.0cm
\hspace*{-2.5cm}
\centerline{\fig{2.8in}{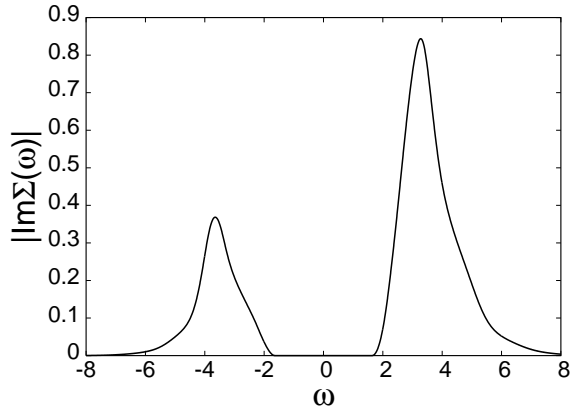}} \vskip 0.5cm \caption{The
imaginary part of self energy $|\Sigma^{\prime\prime}(\omega)|$
for $n=0.5$, $T=0$ and $|U|/t=2.5$ as a function of $\omega$. Note that
$|\Sigma^{\prime\prime}(\omega)|$ becomes non-zero only for
$\omega \ge 3E_g$. For $|U|=2.5t$, $E_g=0.57t$.} \label{imsigma}
\end{center}
\vskip-6mm
\end{figure}

The imaginary part of the (diagonal) self energy is plotted in
Fig.~\ref{imsigma}. It vanishes at low energies ($|\omega| <
3E_g$) since there are no final states available for a scattering
event. In this regime, from eqs.~(\ref{a_def}) and (\ref{g_def})
it follows that \be A(\epsilon,\omega) = \left(\omega+ \epsilon -
\mu + \Sigma(-\omega)\right)\delta\left(D(\epsilon,\omega)\right),
\ee with $D(\epsilon,\omega)=(\omega+\epsilon-\mu+\Sigma(-\omega))
(\omega-\epsilon+\mu-\Sigma(\omega))-S^2(\omega)$, since
$\Sigma(\omega)$ and $S(\omega)$ are now real. The quasiparticle
excitation energies are the two symmetrically placed zeros at
$\omega = \pm E$ of $D$ which is an even function of $\omega$.
Thus the excitation energies are given by \be D(\epsilon,\pm E) =
0. \ee and the residues at the Bogoliubov quasiparticle poles are
then reduced compared to their HFB values and are given by \be
Z_{\pm}(\epsilon) = \lbs \pm E + \epsilon - \mu + \Sigma(\mp
E)\rbs \biggl/\left\vert \f{\partial D}{\partial \omega}(\omega =
\pm E) \right\vert \ee

The imaginary part of the (diagonal) self energy becomes non-zero
for $\omega \ge 3E_g$ as shown in Fig.~\ref{imsigma}. This can be seen to arise from the form of the second order self energy of eq.~(\ref{sigma2}), because in a system with a gap, final states for scattering an injected particle off a particle-hole pair are avaliable only if incident particle has $\omega \ge 3E_g$. We should note, however, that the $3E_g$ value of the threshold is likely an artifact of DMFT/IPT which ignores collective excitations. It is well known \cite{mohit+others3} that in finite dimensions this model has a linearly dispersing sound mode and scattering of a one-particle excitation off such a collective mode should lead to non-zero ${\rm Im}\Sigma(\omega)$ above $E_g$ and not $3E_g$. 

In any case, within IPT, the structure of the self energy leads to the incoherent spectral weight in $A(\epsilon,\omega)$ above three time the gap. The reduction of the coherent quasiparticle weight and the transfer of spectral weight to the incoherent part of the spetral function are thus features related to the $\omega$ dependent self energy and are missing in simple HFB mean field theory. 
 
\begin{figure}
\begin{center}
\hspace*{-5cm} \centerline{\fig{1.5in}{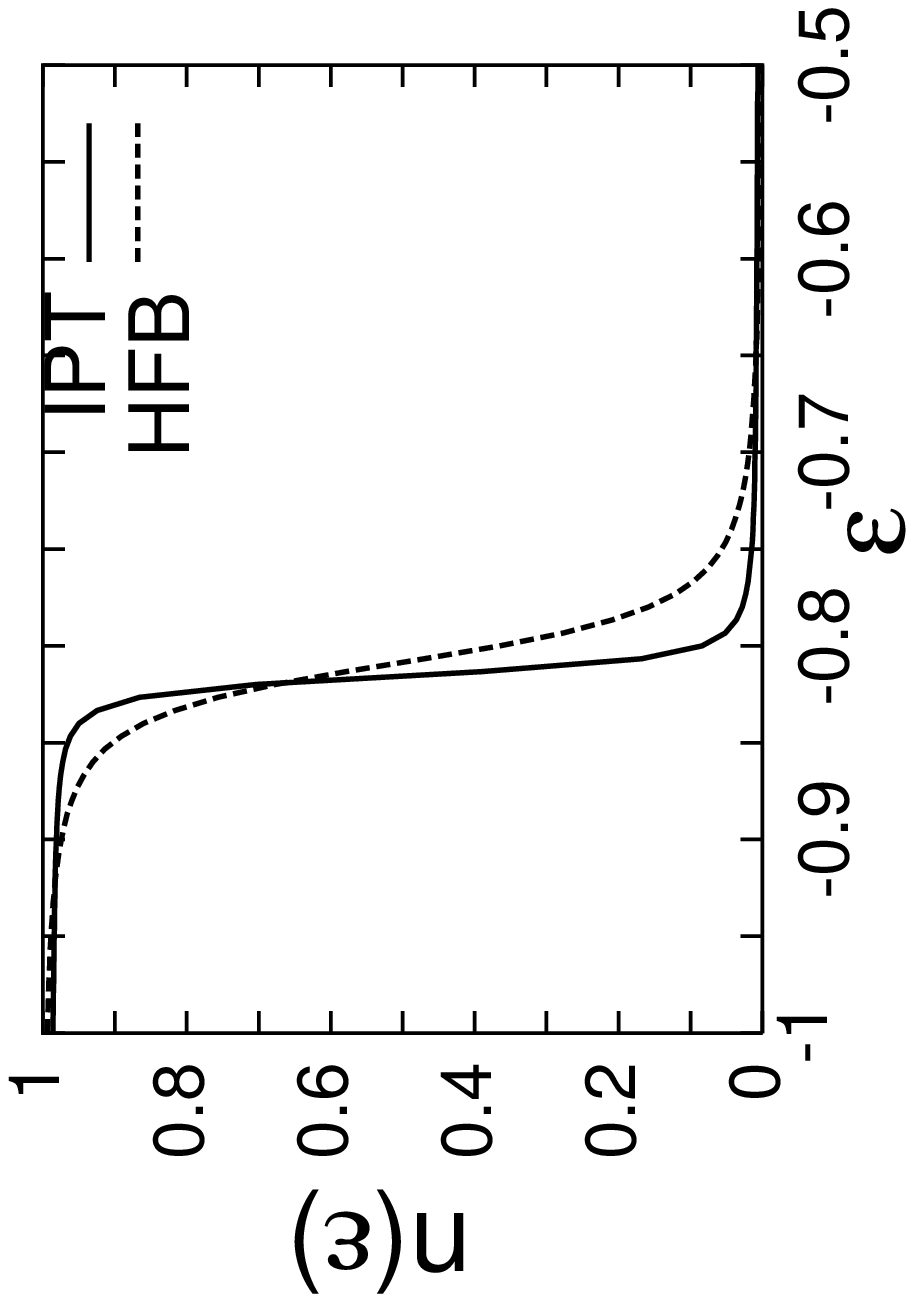}} \vskip-3.85cm
\hspace*{1.8cm} \centerline{\fig{1.5in}{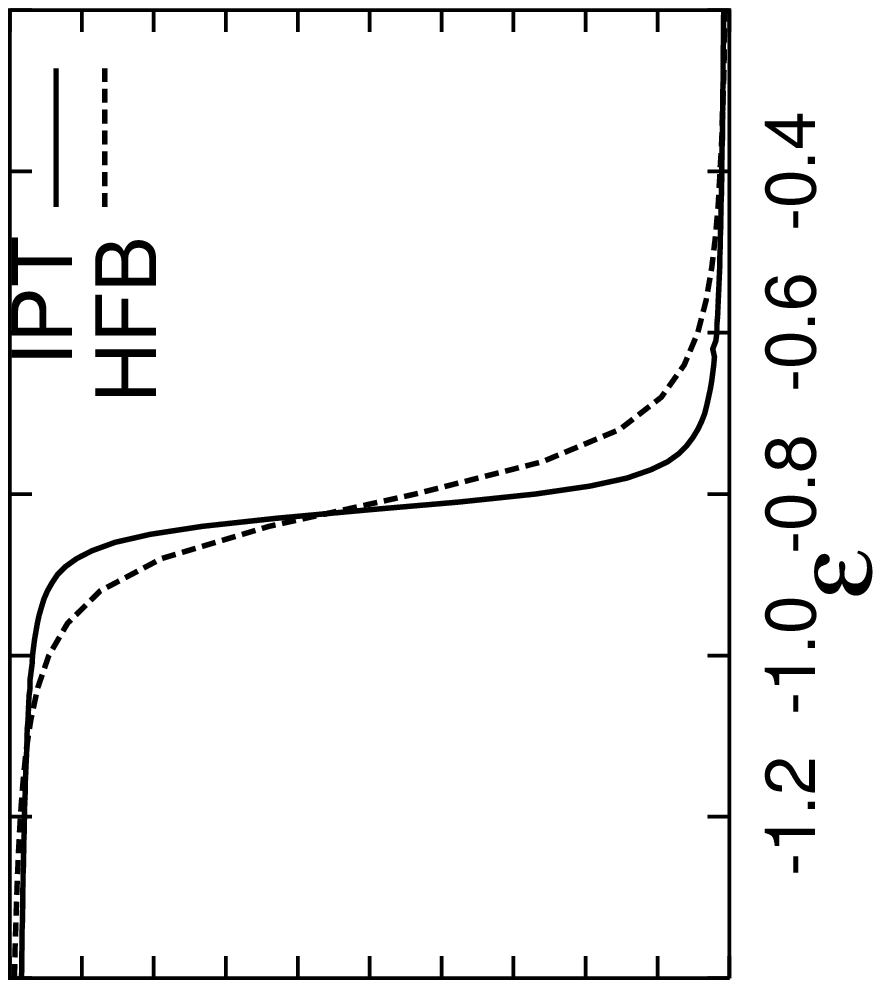}} \vskip1.0cm
\centerline{\fig{2.2in}{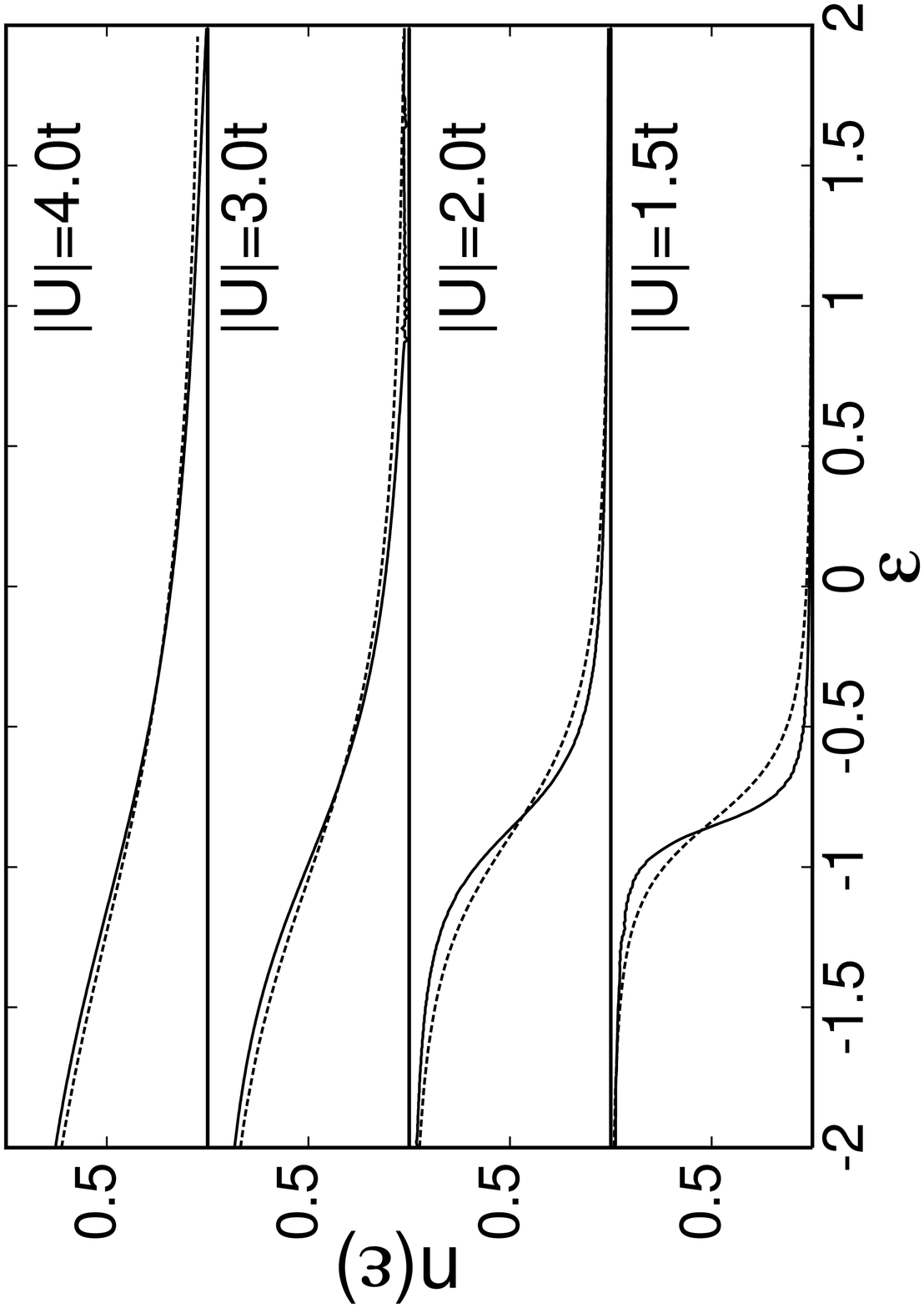}} \vskip 0.5cm \caption{The
occupation probability $n(\epsilon)$ of an energy level $\epsilon$
for $n=0.5$, $T=0$ for various values of $|U|/t$  obtained within
IPT ( full curve) and HFB approximation (dashed curve). Top right
panel shows $n(\epsilon)$ for $|U|=0.75t$ and top left panel shows
$n(\epsilon)$ for $|U|=1.0t$. Note $\epsilon$ scale over which
these are plotted are different. Bottom panel shows $n(\epsilon)$
for $|U|=1.5t,2.0t,3.0t$ and $4.0t$ starting from bottom to top.}
\label{nepslon}
\end{center}
\vskip-6mm
\end{figure}

{\bf Occupation probability:} We next calculate the analog of the
momentum distribution within the DMFT, namely the occupation
probability $n(\epsilon)$ of an energy level $\epsilon$ given by
\be n(\epsilon)=\int_{-\infty}^{0} A(\epsilon,\omega)d \omega. \ee
This is plotted in Fig.~\ref{nepslon} for various values of
$|U|/t$.

Within  the HFB approximation, $n(\epsilon)$ has the following
simple form: \be n(\epsilon)=Z_{-}(\epsilon)=\f{1}{2}\lbr
1-\f{\xi}{E} \rbr \label{nk_HF} \ee It is easy to see that in the
weak coupling limit $n(\epsilon)$ looks like a slightly broadened
Fermi function, dropping from 1 to zero over an energy scale of
order $\Delta$; its width hence increases monotonically with
$|U|/t$. As one begins to form more and more tightly bound pairs,
higher $\epsilon$ states need to be involved in the pairing and
eventually the system becomes non degenerate even at $T=0$ as
already argued from the chemical potential. Note that
$n(\epsilon)$ within IPT is always less rounded than that within
HFB because quantum fluctuations reduce the gap in the single
particle dos relative to HFB value.

We find that the exact sum rule $\int_{-\infty}^{\infty}
n(\epsilon) \rho_0(\epsilon)d\epsilon = n/2$ is satisfied in our
IPT calculation within the estimated numerical errors
($0.4\%-2\%$).

\begin{figure}
\begin{center}
\vskip -2.0cm \hspace*{-2.5cm} \centerline{\fig{2.8in}{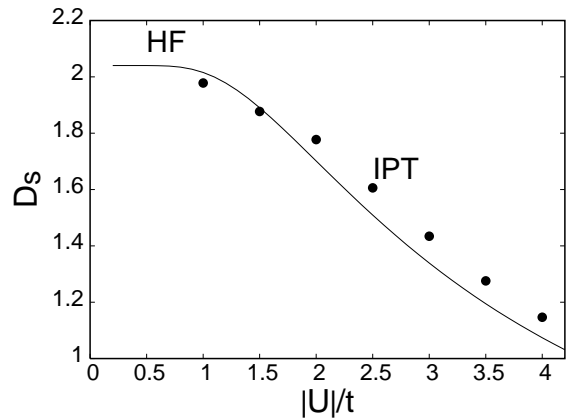}}
\vskip 0.5cm \caption{Upper bound on the superfluid stiffness $D_{s}$ for
$n=0.5$, $T=0$ as a function of the coupling constant $|U|/t$
within IPT (filled circles) and HFB theory (full line).}
\label{superfluid_stiffness}
\end{center}
\vskip-6mm
\end{figure}

{\bf Superfluid stiffness:} We can also estimate an upper bound on
the superfluid stiffness $D_s$ which is the strength of the delta
function in the real part of optical conductivity : \be {\rm
Re}\sigma(\omega) = D_{s} \delta(\omega)+ {\rm
Re}\sigma_{reg}(\omega), \ee The Kubo formula for the superfluid
stiffness~\cite{scalapino} can be written as \be \f{D_s}{\pi} =
-\langle {\mathcal{K}}_{x}\rangle
-Re\Lambda_{T}(q_{x}=0,q_y\rightarrow 0,\omega=0) \ee where the
kinetic energy $-\langle {\mathcal{K}}_{x}\rangle$ is the
diamagnetic response of the system to the vector potential and the
transverse current-current correlation function $\Lambda_{T}$ is
the paramagnetic response. It is easy to see that $\Lambda_{T} \ge
0$, so that $D_{s} \le \pi \vert\langle
{\mathcal{K}}_{x}\rangle\vert$. Thus the the kinetic energy gives
an upper bound to the superfluid stiffness, and in fact we may use
it to provide a rough estimate of $D_s$ (although, as emphasized
in ref.~\cite{paramekanti}, there is no reason to assume in
general that for a lattice model $D_s$ is identical to $\pi
\vert\langle {\mathcal{K}}_{x}\rangle\vert$ even though this
equality holds within simple BCS-HFB theory.) In
Fig.~\ref{superfluid_stiffness} we plot the superfluid stiffness
$D_s$ as a function of the attractive interaction. We find that it
is of order $t$ in weak coupling, but decreases monotonically with
$|U|/t$ reaching $\sim t^{2}/|U|$ in the strong coupling limit,
which reflects the increasing effective mass of the hard core
lattice bosons in the large $|U|$ limit.

\section{6. Conclusions}
In this paper we have studied the crossover from BCS
superconductivity to BEC at $T=0$ in the attractive Hubbard model
using dynamical mean field theory (DMFT), implemented using the
iterated perturbation theory (IPT) scheme. Our main goal was to
explore the DMFT approach in a broken symmetry state, which has
received less attention than the paramagnetic phase, within a
simple, easily implemented, semi-analytic scheme, and to see how
the quantum fluctuations included in this framework modify the
Hartree-Fock-Bogoliubov (HFB) mean field results. For the most
part we found that HFB is qualitatively correct but overestimates
the SC order parameter and energy gap. In the intermediate
coupling regime, the quantitative changes can be quite large. The
frequency dependent self-energy of DMFT leads to the appearance of
incoherent contributions to the single particle spectral function
at energies larger than three times the gap, and the consequent
reduction in the coherent spectral weight in the Bogoliubov
quasiparticle/quasihole poles in the spectrum .

\section{Acknowledgments}
A. G. would like to thank S.R. Hassan and R. Karan for many useful
discussions and acknowledge the hospitality of the Department of
Physics, I.I.Sc., Bangalore, over a period during which a part of
this work was done. M.R.'s work at TIFR was supported in part by
the Department of Science and Technology under a Swarnajayanti
Grant.

\section{Appendix A : Large $\omega$ limit}

In this Appendix we will first determine the large $\omega$ limit
of the self energy and then use it to fix the parameter $\hat{A}$
in our IPT ansatz (\ref{ipt_ansatz}). The $\omega \to \infty$
limit can be obtained by a moment expansion of the Green's
function \be G(\omega^{+})=\f{1}{\omega^{+}}\lbs
M^{(0)}+\f{M^{(1)}}{\omega}+\f{M^{(2)}}{\omega^{2}} + \dots \rbs,
\ee which follows from the spectral representation \be
G(\omega^{+})=\int_{-\infty}^{\infty}
\f{\rho_G(\epsilon)d\epsilon}{\omega^{+}-\epsilon} \ee and the
definition of the $n^{\rm th}$ moment of the density of states
$M^{(n)} = \int_{-\infty}^{\infty}\rho_G(\omega)\omega^{n} d\omega
$.

To evaluate the moments it is useful to go to a Hamiltonian
formulation for single site impurity problem, which requires
introducing auxiliary degrees of freedom to describe the ``bath''.
For the superconducting phase of the negative U Hubbard model, one
possible choice for the impurity Hamiltonian is :
\[H_{imp}=-\mu\sum_{\sigma}c^{\dagger}_{\sigma}c_{\sigma} - |U|n_{\ua}n_{\da} +
\sum_{k\sigma}\epsilon_{k}f^{\dagger}_{k\sigma}f_{k\sigma} \] \be
+\sum_{k \sigma}V_{k} \lbr c^{\dagger}_{\sigma}f_{k
\sigma}+f^{\dagger}_{k \sigma} c_{\sigma}\rbr +D\sum_{k}\lbr
f^{\dagger}_{k\ua}f^{\dagger}_{-k\da} + f_{-k\da}f_{k\ua}\rbr \ee
which describes the impurity $c_{\sigma}$ coupled to
superconducting bath of $f$ fermions. Here $V_{k}$ is the
hybridization parameter which allows fermions to hop between the
bath and the impurity site and the $D$ term represents s-wave
pairing of the $f$'s.

Using the spectral representation, the moments can be written in
terms of commutators ($\left[,\right]$) and anticommutators
($\left\{,\right\}$) involving the $c$'s and the impurity
Hamiltonian: \be \hat{M}_{\alpha\beta}^{(0)}=\langle \left\{
c_{\alpha},c^{\dagger}_{\beta}\right\} \rangle
=\delta_{\alpha\beta} \ee

\be \hat{M}^{(1)}_{\alpha\beta}=\langle \left\{
[c_{\alpha},H_{imp}], c^{\dagger}_{\beta} \right\} \rangle \ee and
\be \hat{M}^{(2)}_{\alpha\beta}=\langle \left\{
[[c_{\alpha},H_{imp}],H_{imp}] c^{\dagger}_{\beta} \right\}
\rangle \ee where $\alpha,\beta=\uparrow,\downarrow$. Explicit
evaluation of these commutators leads to the results \be
\hat{M}^{(1)}=(-\mu-|U|n/2)\hat{\tau}_{z}+\Delta \hat{\tau}_{x}
\ee \be \hat{M}^{(2)}=\lbs \mu^{2}+\f{(2\mu |U|+U^{2})n}{2}\rbs
\hat{\tau}_{0} \ee The Host Green's function $\hat{\mathcal{G}}$
is obtained from the impurity Hamiltonian setting $U=0$, and its
large $\omega$ limit is \be \hat{\mathcal{G}}^{-1}(\omega^{+})
\simeq w^{+}\hat{\tau}_{0}+(\mu-\sum_{k}V_{k}^{2}/\omega)
\hat{\tau}_{z} \ee Using the Dyson equation, the large $\omega$
limit of self energy is then found to be \be
\hat{\Sigma}(\omega^{+})=\hat{\Sigma}_{HFB}+\f{U^{2}n(1-n/2)/2-\Delta^{2}}{\omega^{+}}\hat{\tau}_{0}.
\label{se_largew} \ee

We must now find the large $\omega$ limit of IPT self energy, and
compare it with the exact $\omega \to \infty$ result
(\ref{se_largew}) derived above. Begin with the diagonal component
of $\hat{\Sigma}^{(2)}(\omega)$ given by 
\be
\Sigma^{(2)}(\omega^{+}) =
U^{2}\int_{-\infty}^{\infty}\prod_{i=1}^{3}
d\epsilon_{i}\f{g_{1}(\epsilon_{1},\epsilon_2,\epsilon_3)N(\epsilon_{1},\epsilon_{2},
\epsilon_{3})}{w^{+}-\epsilon_{1}+\epsilon_{2}-\epsilon_{3}} \ee
where \be g_{1}(\epsilon_{1},\epsilon_2,\epsilon_3)=\tilde{\rho}_{11}(\epsilon_{1})
\tilde{\rho}_{22}(\epsilon_{2})\tilde{\rho}_{22}(\epsilon_{3})-\tilde{\rho}_{f}
(\epsilon_{1})\tilde{\rho}_{22}(\epsilon_{2})\tilde{\rho}_{f}(\epsilon_{3})
\ee Here $ \tilde{\rho}_{ii}(\omega)=-1/\pi \mbox{~Im~}
\tilde{\mathcal{G}}_{ii}(\omega^{+})$ with $i=1,2$ and
$\tilde{\rho}_{f}(\omega)=-1/\pi \mbox{~Im~}
\tilde{\mathcal{F}_{0}}(\omega^{+})$. In the large $\omega$ limit
it suffices to keep terms up to order $1/\omega$. We thus get \be
\Sigma^{(2)}(\omega^{+}) \simeq \f{1}{\omega^{+}}\lbs
\f{U^{2}n_{0}}{2}(1-\f{n_{0}}{2})-\Delta_{0}^{2}\rbs \ee Next,
consider the off-diagonal component of
$\hat{\Sigma}^{(2)}(\omega)$ \be S^{(2)}(\omega^{+}) =
U^{2}\int_{-\infty}^{\infty}\prod_{i=1}^{3}
d\epsilon_{i}\f{g_{2}(\epsilon_{1},\epsilon_2,\epsilon_3)N(\epsilon_{1},
\epsilon_{2},\epsilon_{3})}{w^{+}-\epsilon_{1}+\epsilon_{2}-\epsilon_{3}}
\ee where \be g_{2}(\epsilon_{1},\epsilon_2,\epsilon_3)=
\tilde{\rho}_{f}(\epsilon_{1})\tilde{\rho}_{f}(\epsilon_{2})\tilde{\rho}_{f}(\epsilon_{3})
-\tilde{\rho}_{11}(\epsilon_{1})\tilde{\rho}_{f}(\epsilon_{2})\tilde{\rho}_{22}(\epsilon_{3})
\ee It is easy to check that in the large $\omega$ limit $S^{(2)}$
vanishes up to order $1/\omega$.

Comparing the large $\omega$ limits of the IPT ansatz
$\hat{\Sigma}_{HFB} + \hat{A}\hat{\Sigma}^{(2)}(\omega)$ and the
exact self energy (\ref{se_largew}), we find \be \hat{A}=\lbs
\f{U^{2}n_{0}}{2}\lbr 1-\f{n_{0}}{2}\rbr -\Delta_{0}^{2} \rbs^{-1}
\lbs\f{U^{2}n}{2}\lbr 1-\f{n}{2}\rbr -\Delta^{2}\rbs
\hat{\tau}_{0} \ee

\section{Appendix B : Atomic limit}
In this Appendix first we first solve the attractive Hubbard model
exactly in the atomic limit $t/U=0$ and then show that our IPT
ansatz for self energy is exact in this limit. In the atomic limit
one can drop the hopping term in the Hubbard Hamiltonian (and also
the hybridization term in the impurity Hamiltonian) so that \be H=
-|U| n_{\ua}n_{\da}-\mu(n_{\ua}+n_{\da}) \ee The various sites
decouple and so we have dropped the site label. The four states
are $|0\rangle, |\ua\rangle, |\da \rangle$ and $|\ua\da\rangle$
with corresponding energies $0,-\mu,-\mu$ and $-2\mu -|U|$.

To study the broken symmetry phase we introduce a pairing field h,
\be
 H= -|U|n_{\ua}n_{\da}-\mu(n_{\ua}+n_{\da})-h(c^{\dagger}_{\ua}c^{\dagger}_{\da} + c_{\da}c_{\ua})
\ee and finally take the $h \to 0$ limit. The $T=0$ equations for
the filling factor and order parameter are given by \be
n=\f{2}{1+(h/\lambda)^{2}} \mbox{~and~}\Delta=
\f{-|U|h/\lambda}{1+(h/\lambda)^{2}} \ee where $\lambda
=(-(2\mu+|U|)-\sqrt{(2\mu+|U|)^{2}+4h^{2}})/2$ is the lowest
eigenvalue of the Hamiltonian. We thus find that in the
$h\rightarrow 0$ limit we get the atomic limit solution:
$\mu=-|U|/2$ for any filling $n$ and $\Delta=|U|\sqrt{n(2-n)}/2$.

Now consider the HFB equations in the atomic limit at $T=0$ \be n
= 2\lbs 1+\f{\mu+|U|n/2}{\sqrt{(\mu+|U|n/2)^{2}+\Delta^{2}}}\rbs
\ee and \be \f{1}{|U|} =
\f{1}{2\sqrt{(\mu+|U|n/2)^{2}+\Delta^{2}}} \ee The solution of
these self consistent equations is $\mu=-|U|/2$ and
$\Delta=|U|\sqrt{n(2-n)}/2$, which shows that HFB theory is exact
in the atomic limit at zero temperature.

Finally we will show that the second order self energy vanishes in
the atomic limit. The Hartree corrected Host Green's function \be
\hat{\tilde{\mathcal{G}}_{}}^{-1}(w^{+})=\hat{\mathcal{G}}_{}^{-1}(w^{+})
- \hat{\Sigma}_{HFB} \ee reduces in the atomic limit to \be
\hat{\tilde{\mathcal{G}}_{}}^{-1}(w^{+})=
w^{+}\hat{\tau}_{0}+(\mu+|U|n/2)\hat{\tau}_{z}+\Delta\hat{\tau}_{x}
\ee It can be checked that the full Green's function in the atomic
limit is identical to this, which means that
$\hat{\Sigma}^{(2)}(\omega)$ vanishes. Alternatively one can
calculate the density of states corresponding to
$\hat{\tilde{\mathcal{G}}_{0}}$, which are given by \be
\tilde{\rho}_{11}(w)=(1-n/2)\delta(w-|U|/2)+(n/2)\delta(w+|U|/2)=\tilde{\rho}_{22}(-\omega)
\ee and \be
\tilde{\rho}_{f}(w)=-\Delta/|U|(\delta(w-|U|/2)-\delta(w+|U|/2))
\ee and substitute these in the expression for the
$\hat{\Sigma}^{(2)}(\omega)$ and check that all the components of
second order self energy matrix vanish in the atomic limit.


\begin{thebibliography}{99}
\bibitem{lt-note}
Note that we do not impose the Luttinger Integral Theorem (also
refereed to as the Luttinger sum rule). While its imposition is of 
importance in metallic systems to ensure Fermi liquid
properties \cite{georges,kk}, we believe that its importance is
not as crucial in gapped systems such as the ones under discussion.
\bibitem{ipt-hfb-note}
We note that in the presence of the IPT contributions to the self
energy, the values of $\mu$ and $\Delta$, obtained from eqs. 
(\ref{delta},\ref{no_eqn}), are different from the values that 
arise in pure HFB.
\end{thebibliography}
\end{document}